# BEAM EXTRACTION FROM THE RECYCLER RING TO P1 LINE AT FERMILAB*

M. Xiao†, D. Capista, P. Adams, D. Morris, M.J. Yang, K. Hazewood

Fermilab, Batavia, IL 60540, USA

*Abstract*

The transfer line for beam extraction from the Recycler ring to P1 line provides a way to deliver 8 GeV kinetic energy protons from the Booster to the Delivery ring, via the Recycler, using existing beam transport lines, and without the need for new civil construction. It was designed in 2012. The kicker magnets at RR520 and the lambertson magnet at RR522 in the RR were installed in 2014 Summer Shutdown, the elements of RR to P1 Stub (permanent quads, trim quads, correctors, BPMs, the toroid at 703 and vertical bending dipole at V703 (ADCW) were installed in 2015 Summer Shutdown. On Tuesday, June 21, 2016, beam line from the Recycler Ring to P1 line was commissioned. The detailed results will be presented in this report.

## INTRODUCTION

In the post-Nova era at Fermilab complex, shown in Fig. 1, the protons are directly transported from the Booster ring to the Recycler Ring (RR) rather than the Main Injector (MI) [1]. For Mu2e and g-2 projects, a new beamline, shown in RED in Fig. 1, was designed in 2012 [2] and completed in the installation in 2015. This new beamline provides a way to deliver 8 GeV kinetic energy protons from the Booster to the Delivery ring, via the Recycler, using existing beam transport lines (P1, P2 line), and without the need for new civil construction. Fig. 2 presents the schematic layout of the transfer line from the Recycler Ring to the P1, P2 line. Shown in Blue in Fig. 2 are the kicker magnets RRKICK and the lambertson magnet RRLAM in the RR, which were installed in 2014 Summer Shutdown.

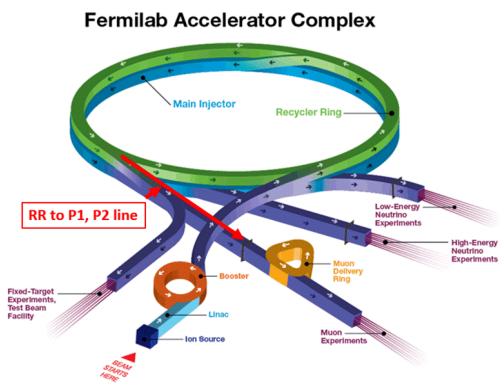

Figure 1: Fermilab Complex.

___________________
*Work supported by U.S. Department of Energy under contract No. DE-AC02-76CH03000.
† meiqin@fnal.gov

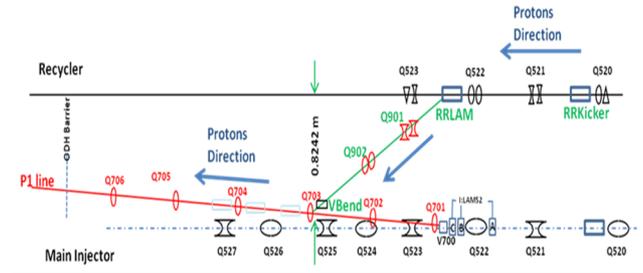

Figure 2: Schematic layout of the transfer line.

The Stub from the RR to P1 line (shown in green in Fig.2), including permanent quads, trim quads, correctors, BPMs, the toroid at 703 and vertical bending dipole at V703 (ADCW), was installed in 2015 Summer Shutdown. A new 2.5MHz RF system, which will be used to rebunch a $4 \times 10^{12}$ of protons from Booster into 4 bunches of $10^{12}$ protons in the Recycler, is being installed during 2016 Summer Shutdown. The commissioning for the beam extraction was done at the end of June 2016, the protons were extracted successfully from the Recycler ring to P2 line. The detailed results will be presented in this report.

## DEVICES AND PARAMETER SCANS

The main devices and their best values for RR to P1 line extraction are shown in Fig. 3. The parameter R: KPS5A is the voltage for the kicker magnets RRKICK, R: LAM52 is the current for the lambertson magnet and R: V703 is the current for the vertical bending .Trim dipoles R:VT701 and R:HT702, trim quads R:QT701 and R:QT702 were not used at first run. These best values left in the parameter page R65 are based on the parameter scan for R:KPS5A and R:LAM52.

Figure 3: Best values of the devices for beam extraction.

The scans were done with the changes of the values of R:KPS5A and R:LAM52, by monitoring 3 loss monitor R: LI522F, R: LI522E and R: LI522G , which are located up, middle and downstream of the lambertson magnet, the results are shown in Fig.4 and Fig. 5.

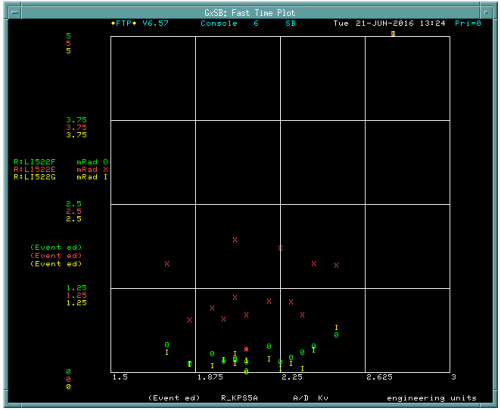

Figure 4: Scan for kicker magnet strength.

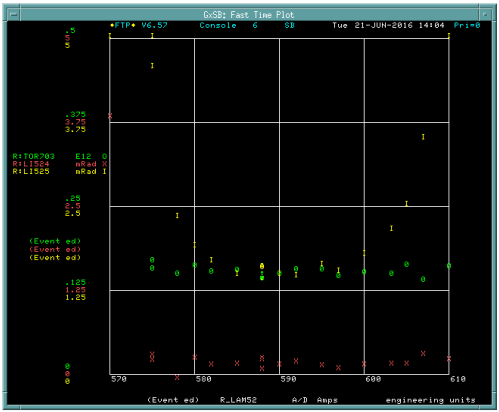

Figure 5: Scan for Lambertson magnet strength.

One bumps scan with trim dipoles was done on June 23, 2016. It made ramps for R: HT702 and R:VT701 and changed in 1A increments, as shown in Figure 6, and 7.

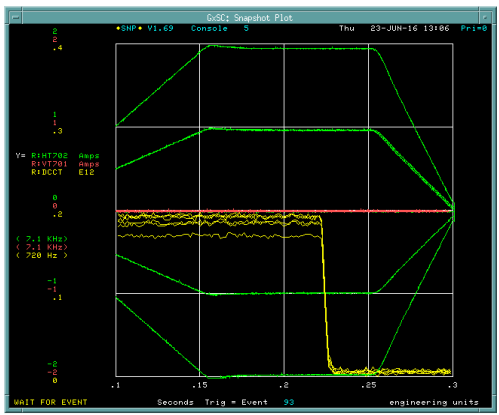

Figure 6: Scan for kicker magnet strength.

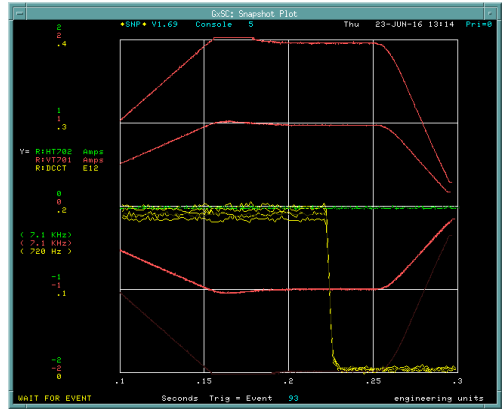

Figure 7: Scan for kicker magnet strength.

## BEAM EXTRACTION AND THE PROFILES OBTAINED BY MULTIWARES

Due to not yet available of the 2.5MHz RF system in the RR at the time of commissioning, 15 bunches of the proton beams in 53MHz beam structure, each accumulated in 4-6 Boosters turns, was extracted from the RR to P1, P2 line. The intensity of the protons is varied from 1.4 to $1.8E10^{11}$, and the estimated emittance is about 12 π mm·mrad. Indicated in Fig. 8 are the value of the Recycler DCCT (green), and those on Toroid 703 in the RR to P1 Stub (Red), on Toroid 714 in the P1 line (Yellow) as well as Toroid 716 in the P2 line (Cyan). The vertical scale for R:DCCT is 1.0E12, and 0.5E12 for the rest of 3 Toroid's. At the end of scan, about 1.7E11 of the beam shown on Toroid 716, the overall extraction efficiency is above 90%.

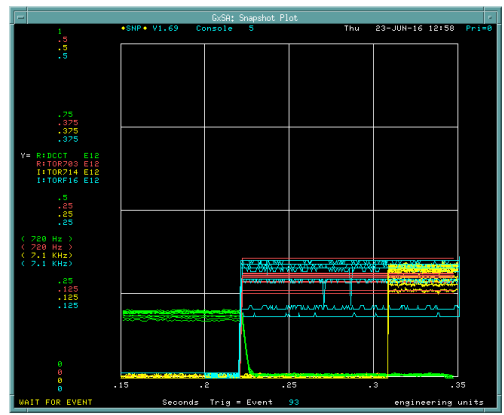

Figure 8: Beam intensity shown during the extraction.

Horizontal and vertical BPMs along the RR to P1 line Stub, P1 and P2 line are triggered and the readings are shown in Fig. 9 and Fig. 10.

## BEAM PROFILES AND OPTICS MATCH

Figure 2 shows the beam profiles in both horizontal and vertical planes obtained by the multiwares at F11, F12, F13 and F17 respectively. Based on the the estimated emittances of the beam and the beta-function obtained by optics match, we calculated the beam size at each locations of the multiwares, the results are listed in Table 1.

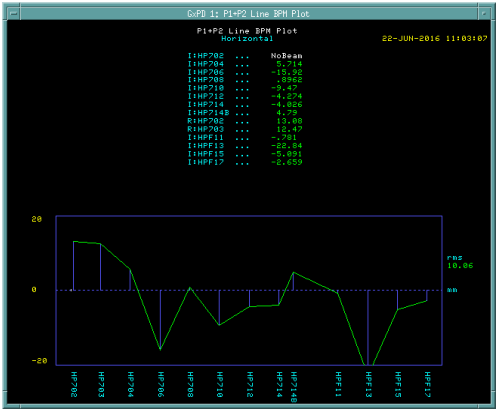

Figure 9. Beam intensity shown during the extraction

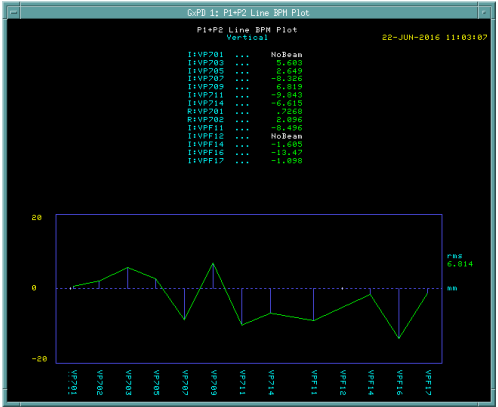

Figure 10. Beam intensity shown during the extraction

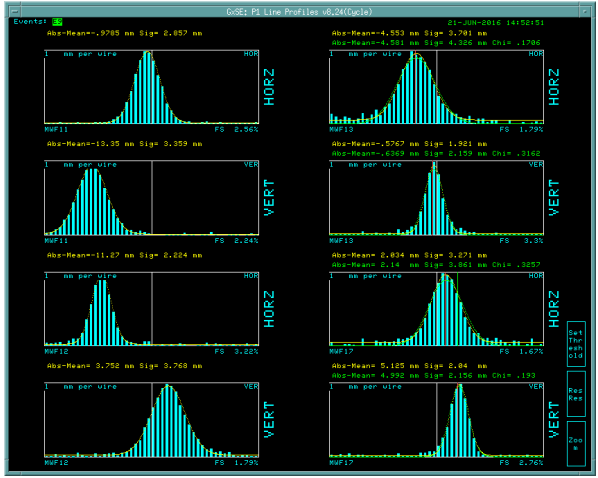

Figure 11. Beam intensity shown during the extraction

The beam sizes between calculated and measured are not far away from each other. However, we found that the requested settings of the current for Quad Q710 was 10 times less due to misuse of a transfer constant when it was converted from the magnetic strength. Actually, the optics match was done from the end of lambertson magnet in the Recycler ring to the end of P2 line. The optimized match results request that

$$\begin{cases} K_1QR\,701 = -0.071\,(1/m^2) \\ K_1QR\,702 = +0.090\,(1/m^2) \end{cases} \quad \begin{cases} K_1Q3 = -0.0414\,(1/m^2) \\ K_1Q10 = +0.0214\,(1/m^2) \end{cases}$$

Table 1: beam sizes measured and calculated

|  | F11 | F12 | F13 | F17 |
|---|---|---|---|---|
| $\beta_{x\,(m)}$ | 25.5 | 11.61 | 94.36 | 103.61 |
| $\beta_{y\,(M)}$ | 53.08 | 124.13 | 29.00 | 27.34 |
| $\sigma_{x\,(calculated)}$ (mm) | 2.32 | 1.56 | 4.46 | 4.66 |
| $\sigma_{x\,(measured)}$ (mm) | 2.857 | 2.224 | 3.701 | 3.271 |
| $\sigma_{y(calculated)}$ (mm) | 3.35 | 5.12 | 2.47 | 2.40 |
| $\sigma_{y(measured)}$ (mm) | 3.359 | 3.760 | 1.921 | 2.04 |

$K_1QR701$ and $k_1QR702$ are the quadrupole strengths of the permanent magnets QR701A&B and QR702A&B, and the measured strengths meet the requests. $K_1Q3$ is the quadrupole strength of Q703, Q704,… Q709 in a FODO lattice section. IQ703T is the current of the circuit for all these quads, positive sign for the focusing quads, negative sign for defocusing quads. $K_1Q10$ is the quadrupole strength for Q710. Based on the transfer constants: TF_Q703=0.12228 and TF_Q710=1.5109, the currents requested should be as follows

$$IQ703T = 214.47\,Amps$$
$$IQ710T = 12.80\,Amps$$

But it was requested for 1.25Amps. Actual settings was 2.0 Amps due to the device limit.

## CONCLUSION

The beam extraction from the Recycler Ring to the P1, P2 line for Mu2e and G-2 project was successfully commissioned. The transfer efficiency is above 90% for 1.85E11 of the protons in the RR. All the devices and the instrumentations worked well. Further commissioning will be done after the installation of the new 2.5MHz RF system in this summer shutdown in 2016.

## ACKNOWLEDGEMENT

The commissioning was done by the people of the whole MI/Recycler group and the Muon group. We would like to thank all the people involved, also people from EE Suport and operators in the Accelerator Division at Fermilab.